\documentclass[preprint,epsfig,pre]{revtex4}

\usepackage{diagbox}
\usepackage{graphicx}
\usepackage{epstopdf}
\usepackage{amssymb}
\usepackage{amsmath}
\usepackage{latexsym}
\usepackage{color}
\usepackage[utf8]{inputenc}
\usepackage[T1]{fontenc}

\begin{document}

\renewcommand{\thefootnote}{\fnsymbol{footnote}}
\renewcommand{\theequation}{\arabic{section}.\arabic{equation}}

\title{Effect of topology on the statics and dynamics of a polymer chain at the fluid-fluid interface: a molecular dynamics simulation study}

\author{Jenis Thongam} 
\author{Lenin S.~Shagolsem}
\email{slenin2001@gmail.com}

\affiliation{Department of Physics, National Institute of Technology Manipur, Imphal - 795004, India}
\date{\today}

\begin{abstract}
The effect of chain topology on the statics and dynamics of chain at the interface of two immiscible fluids is studied by means of molecular dynamics simulations. For three topologically different chains, namely, linear, ring, and trefoil-knot of same molecular weight the effect of varying both polymer--fluid and fluid--fluid interaction nature on the width of the fluid interface, chain conformation, shape, and chain dynamics. For sharp-interface binary-fluid system, the interface width is insensitive to both topology and polymer-fluid interaction nature, while a weak non-monotonic variation is seen for weak-interface system. Chain extension normal to the interface plane is significantly affected by the topology with trefoil-knot chain, due to the additional constraint, has the largest value compared to both linear and ring polymers. Instantaneous shapes are also quantified through shape parameters. Furthermore, it is observed that the qualitative behavior of center of mass mean-square displacement (MSD) is independent of topology, i.e., all the chain types show same diffusion exponent $\alpha~(\approx 1)$. However, the self-diffusion constant depends on the topology and it is largest for trefoil-knot chain. An interesting observation pertaining the early time behavior of monomeric-MSD is that, within the sub-diffusive regime, the values of $\alpha$ for different parameters (independent of topology) are grouped into two distinct ranges (0.52--0.59 and 0.62--0.67) which is related to the different chain conformation for polymer-fluid interaction range below and above a threshold value equal to that of the self-interaction of the pure fluid phase. 
\end{abstract}

\maketitle

\section{Introduction}
\label{sec: introduction}

Understanding the behavior of polymer at the interface is of scientific and technological interest because of its potential applications, e.g., as surfactants and colloidal stabilization among zoo of aplications. One of the important factor which made it possible is the polymer's ability to change conformation when exposed to different environments (e.g.~poor solvents, water-air or water-oil interfaces) so as to minimize their free energy.\cite{gennes1979,doi1988,pereira2008} For instance, when proteins which consists of hydrophobic and hydrophilic parts, come in contact with a biological membrane immersed in water environments, they get adsorbed at the lipid-water interface and each residues have the tendency to reside in its favourable solvent.\cite{ermoshkin2002, rother1998} Since many biological macromolecules contain both hydrophilic and hydrophobic groups, it is reasonable to use the liquid-liquid interface when manipulations of such molecules are required. For example, the extraction of ssDNA is carried out by using water-phenol interface in a method called PERT (phenol emulsion reassociation technique),\cite{goldar2004, kohne1977, miller1995, bruzel2006} were the extraction is caused by the amphiphilic composition of macromolecules built of hydrophilic sugar-phosphate background and hydrophobic bases. \\

On the other hand, homopolymers (which is not amphiphilic) can get attracted to the interface where it helps in lowering the interface tension by screening the unfavorable contacts between solvent molecules of different kinds and thus act as emulsion stabilization agent.\cite{halperin1986} 
For example, microgels adsorbed at a liquid-liquid interface reduces the interfacial tension by reducing the number of unfavourable contacts between the immiscible molecules.\cite{ramsden1903, pickering1907} Also, in the water-oil interface, microgels substitute the water-oil contacts by polymer-water and polymer-oil contacts.\cite{artem2016} This reduction in interfacial tension induces the microgels to spread over the interface all the while maintaining their spherical shape in each of the liquids.\cite{destribats2011, schmidt2011} And since the molecules of both liquids can penetrate the adsorbed microgels, parts of the microgels exposed to the different liquids are able to swell or collapse differently depending on the solvent quality of the liquids and this explains their surface activity.\cite{artem2016} Yet another system of polymer at interface is that of polymer confined in thin film geometry and the effect it has on the conformation of the chain both perpendicular and parallel to the interface.\cite{brandon2017}  
Interesting properties such as formation of compact objects with fractal contours, dynamics and disentanglement, algebraic displacement correlation, are observed in thin and two-dimensional polymer films.\cite{carmesin1990,meyer2007,meyer2009,wittmer2010} On the other hand, topological properties of macromolecules are also of interest in various aspects.\cite{richter2015,polymeropoulos2017} For instance, in material science, polymer topology plays an importnat role in tuning the physical properties (e.g., strength, toughness, glass transition, and mechanical response) leading to interesting applications ranging from coating to drug and gene delivery.\cite{Kapnistos2008,Gillies2005,Gao2004} Also, in biological systems, the protein folding is related to the topological complexity of the native state.\cite{kevin1998} 
In the current work, of all the possible chain topology, (in addition to linear) we are interested in ring and trefoil-knot topology which are observed in proteins and DNA.\cite{mcleish2002,semlyen1994,sumners2011,millett2018} Prokaryotic genomic DNAs and many viral DNAs are circular, i.e.~without free ends, and knots in such a closed DNAs can form during replication and transcription. Furthermore, it has shown experimentally that knots in protiens occured in denatured state, and that the required denaturation rate to knot depends on their structural complexity. \cite{sumners2011,millett2018} See reference~\cite{halverson2014} and \cite{hofmann2015} for a review on the role topological constraint in genome folding and the role of loops on order of eukaryotes and prokaryotes, respectively. \\

In this work, motivated by the importance and relevance of polymer topology in diverse area ranging from material science to biological systems, we attempt to understand the effect of topology on the static and dynamic behavior of a model polymer chain at the fluid-fluid interface by means of computer simulations. In particular, the behavior of three topologically different chains, namely, linear, ring and trefoil trapped at the liquid-liquid interface is investigated with respect to how their equilibrium static properties such as size, shape, and dynamics repond to changing polymer-fluid and fluid-fluid interactions behavior. A recent experimental work related to the current study is reported by Bergfreund and coworkers, where they have investigated globular protein layer assembly, including structural transition and network formation, at oil-water interface.\cite{bergfreund2021} They observed a significant effect of oil polarity on the degree of interfacial protein unfolding, protein location, and network formation. \\

The remainder of our paper is organized as follows: In section~\ref{sec: model-simulation-details}, we present the model and the simulation details. Effect on the fluid-fluid interface width due to the presence of topologically different polymers and monomer distribution in the interface region are discussed in section~\ref{sec: interface width}. And in section~\ref{sec: chain size and shape} the average chain size and instantaneous shapes are studied, respectively, which is followed by the study of chain dynamics along the interface plane in section~\ref{sec: dynamics}. Finally, we conclude with a summary of the results and discuss in section~\ref{sec: conclusion}.



\section{Model and simulation details}
\label{sec: model-simulation-details}

The system under study consists of a coarse-grained polymer chain confined at the fluid-fluid interface of a symmetric binary-fluid system. The polymer chain is represented by generic bead-spring model (Kremer-Grest model).\cite{kremer_JCP_92} And fluid, a phase separating binary mixture (50:50) of fluid-A and fluid-B, is also represented by coarse-grained particles. All the pairwise interactions in the system is modeled via Lennard-Jones (LJ) potential defined as
\begin{equation}
U_{\tiny{_{\rm LJ}}}(r) = 4\epsilon\left[(\sigma/r)^{12}-(\sigma/r)^{6}\right]~,
\label{eqn: LJ-potential}
\end{equation}
with $r$ the separation between a pair of particles, $\epsilon$ being the depth of the potential, and $\sigma$ the effective coarse-grained monomer diameter. We have chosen the size of the fluid particles to be same as that of the monomer. The LJ potential is truncated and shifted to zero at a cut-off distance $r_c$ (in units of $\sigma$) which is different for different pairs as summarized in table~\ref{table: cut-off-radii}. Here, we fix the value of $\sigma$ as well as $\epsilon$ to be unity. For the binary-fluid, AA and BB self-interaction strength are the same (whose value is fixed by setting the potential cut-off distance $r_c=r_{c}^*=1.5\sigma$), but there is an additional repulsive energy between the adjacent particles of fluid-A and fluid-B (which we introduce in the system by setting the interaction cut-off distance in the range $r_{c1}=1.12-1.35$). This choice of $r_{c1}$ values allow the fluid-fluid interface to vary from having a sharp-interface (for $r_{c1}=1.12$) to a weak-interface (for $r_{c1}=1.35$). And for the polymer-fluid interaction the cut-off distance $r_{c2}$ is varied in the range $1.12-2.0$ thereby changing the nature of interaction from being repulsive to attractive. (Alternatively, the polymer-fluid attraction strength could also be varied by changing $\epsilon$ rather than $r_{c2}$.) So the polymer-fluid interaction strength changes from below to above that of the fluid-AA (or BB) self-interaction. \\
On the other hand, the monomer connectivity along the chain is assured via finitely extensible, nonlinear, elastic (FENE) springs~\cite{grest_kramer_PRA_33} represented by the potential, 
\begin{equation}
U_{\tiny{_\text{FENE}}} = \left\{
\begin{array}{l l}
-\frac{kr_0^2}{2}\ln\left[1-\left(r/r_0\right)^{2}\right]~, & \quad r<r_0 \\ 
\infty~, & \quad r\ge r_0
\end{array} \right.
\label{eqn: fene-potential}
\end{equation}
where $r$ is the separation of neighboring monomers in a chain. The spring constant $k=30\epsilon/\sigma^2$ and the maximum extension between two consecutive monomers along the chain $r_0=1.5\sigma$. The above values of $k$ and $r_0$ ensure that the chains avoid bond crossing and very high frequency modes.\cite{kremer_JCP_92} \

\begin{table}[h]
\caption{Pair-wise interaction range among the species. In the table, the cut-off distance for the self-interaction of the binary-fluid is denoted by $r_{c}^*$.}
\centering
\begin{tabular}{@{}lll@{}}
	\hline
	Interaction between & Cut-off radius $r_c/\sigma$ & Nature of interaction \\
	\hline
	monomer -- monomer			& 1.12       	 & Repulsive        \\
	monomer -- fluid-A/B  		& 1.12 -- 2.00   & Repulsive -- Attractive        \\
	fluid-A -- fluid-A        	& $r_{c}^*$ (= 1.50)       	 & Attractive        \\
	fluid-B -- fluid-B        	& $r_{c}^*$ (= 1.50)       	 & Attractive        \\
	fluid-A -- fluid-B        	& 1.12 -- 1.35   & Repulsive -- weakly-attractive   \\
	\hline
\end{tabular}
\label{table: cut-off-radii}
\end{table}

To investigate the static and dynamic properties we perform constant NVT molecular dynamics (MD) simulations using Langevin dynamics\cite{allen-book,frenkel-smit-book} where the equations of motion are given by  
\begin{equation}
m_i \frac{d^2 {\bf r}_i}{dt^2} + \gamma \frac{d {\bf r}_i}{dt} = -\frac{\partial U}{\partial {\bf r}_i} + {\bf f}_i(t)~,
\label{eqn: langevin}
\end{equation}
with ${\bf r}_i$ and $m_i$ the position and mass of particle $i$, respectively, $\gamma$ the friction coefficient which is taken to be the same for all particles, $U=U_{\tiny{_{\rm LJ}}}+U_{\tiny{_\text{FENE}}}$ is the potential acting on monomer $i$ of the polymer chain, and ${\bf f}_i$ are random external forces which follows the relations: $\left\langle {{\bf f}_i}(t)\right\rangle=0$ and $\left\langle {{\bf f}_i}^\alpha(t){{\bf f}_j}^\beta(t')\right\rangle=2\gamma m_i k_{_B}T\delta_{ij}\delta_{\alpha\beta}\delta(t-t')$ where $\alpha$ and $\beta$ denotes the Cartesian components. 
The friction coefficient can be written as $\gamma=1/\tau_d$, with $\tau_d$ the characteristic viscous damping time which we fixed at $50$ and it determines the transition from inertial to overdamped motion in the dilute system limit. The chosen value of $\gamma$ gives correct thermalization for our study. The equations of motion are integrated using velocity-Verlet scheme with a time step of $\delta t_0 = 0.0025 \tau_{_{\rm LJ}}$. All the physical quantities are expressed in terms of LJ reduced units where $\sigma$ and $\epsilon$ are the basic length and energy scales respectively. The reduced temperature $T$ and time $t$ are defined as $T = k_BT_0/\epsilon$ and $t = t_0/\tau_{_{\rm LJ}}$, where $\tau_{_{\rm LJ}}=\sigma\sqrt{m/\epsilon}$ represents the LJ time unit, and $k_B$, $T_0$, $t_0$, and $m$ are the Boltzmann constant, absolute temperature, real time, and mass, respectively. \\

In this study, all the three different chains (linear, ring, trefoil-knot) have same numbers of monomers per chain, $N=128$. In total, there are $N_{\rm tot}= 43,264~(\rm polymer+fluid)$ particles confined in a simulation box of dimensions $L_x = L_y = 36.35$, and $L_z=37.35$. Thus, the number density of the system $\rho^{\ast}\approx0.88~(=N_{\rm tot}/{\rm Vol.})$.
In order to restrict the position of fluid-fluid interface during the sumulations, we confine the system with two virtual walls (through 93-LJ potential) located at $z=-18.675\sigma$ and $z=18.675\sigma$. Thus, the simulation box is non-periodic along the Z-direction and periodic along the X- and Y- directions. The energy of particle-wall interaction is given by $U_{\rm pw} = \epsilon_{\rm pw}\left[ \frac{2}{15}(\sigma_{\rm pw}/r)^9-(\sigma_{\rm pw}/r)^3 \right]$ for $r<r_c$, with $\epsilon_{\rm pw}=1,~\sigma_{\rm pw}=1$, and interaction cut-off distance $r_c$ at potential minimum. The systems are relaxed at $T=1.0$ (at which binary-fluid macrophase separate forming A-rich and B-rich regions) for about $5\times10^6$ MD steps followed by production runs of $20\times10^6$ MD steps at the same temperature where various measurements are perform.
All the simulations are carried out using open source simulation package LAMMPS.\cite{lammps} 


\begin{figure}[ht]
	\begin{center}
		\includegraphics[trim=1cm 3.5cm 4cm 2cm, clip=true,width=0.9\textwidth]{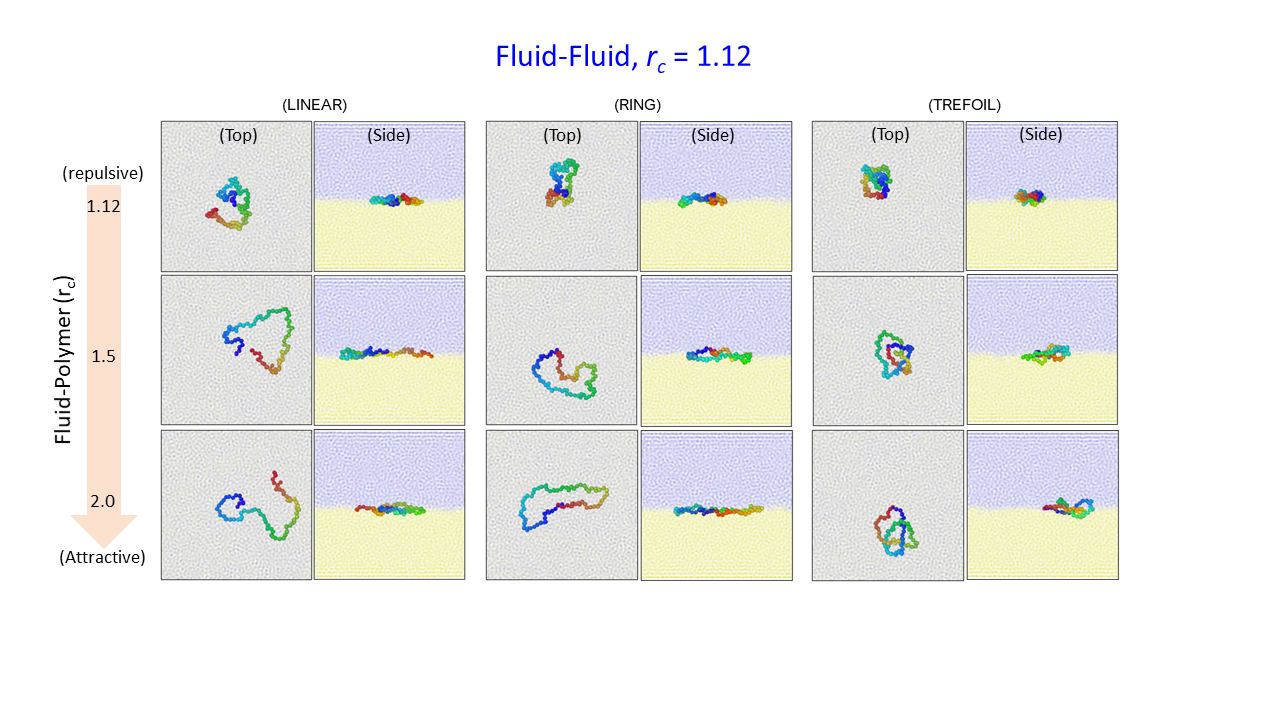}
		\caption{MD simulation snaphots showing (top and side view of) typical instantaneous configuration of the linear, ring and trefoil chains located at the interface of two immiscible fluids (case of sharp interface). Fluids are shown in smeared out blue and yellow regions to clearly see the polymer. In order to identify easily the different regions of the chain, monomers are give continuous color scheme (blue to red of the visible spectrum).}
		\label{fig: simulation-snapshot}
	\end{center} 
\end{figure}

\section{Results}

Typical equilibrium configurations of linear, ring, and trefoil chains at the fluid-fluid interface obtained from the simulations is shown in figure~\ref{fig: simulation-snapshot} for the case of strongly segregated binary fluid having sharp interface and polymer-fluid interaction ranging from purely repulsive to attactive. From the figure it is clear that the chains adopts a compact structure when the polymer-fluid interaction is repulsive and as we gradually changes the interaction nature to attractive the chain starts swelling in the lateral direction. As one would have expected, the lateral extension is observed to be maximum for linear chain followed by ring and trefoil. And extension in the direction perpendicular to interface plan is least (or negligible) for linear and ring polymers, whereas for trefoil the topological constrain, i.e. presence of knot, make the extension appreciable in the perpendicular direction. (Equilibrium configurations of the chains for relatively weak binary-fluid interface in included as supplementary figures.) 
Thus, the topologically different chains under consideration respond differently upon varying polymer-fluid interaction and as a consequence may affect the interface width differently even at this very low polymer concentration. In the following, we investigate the interplay between the chain topology and interface width, in particular, we characterize the fluid-fluid interface width and also check the monomer distribution in this narrow interface region. 

\subsection{Interface width and monomer distribution}
\label{sec: interface width}

\begin{figure}[ht]
	\begin{center}
		\includegraphics[width=0.9\textwidth]{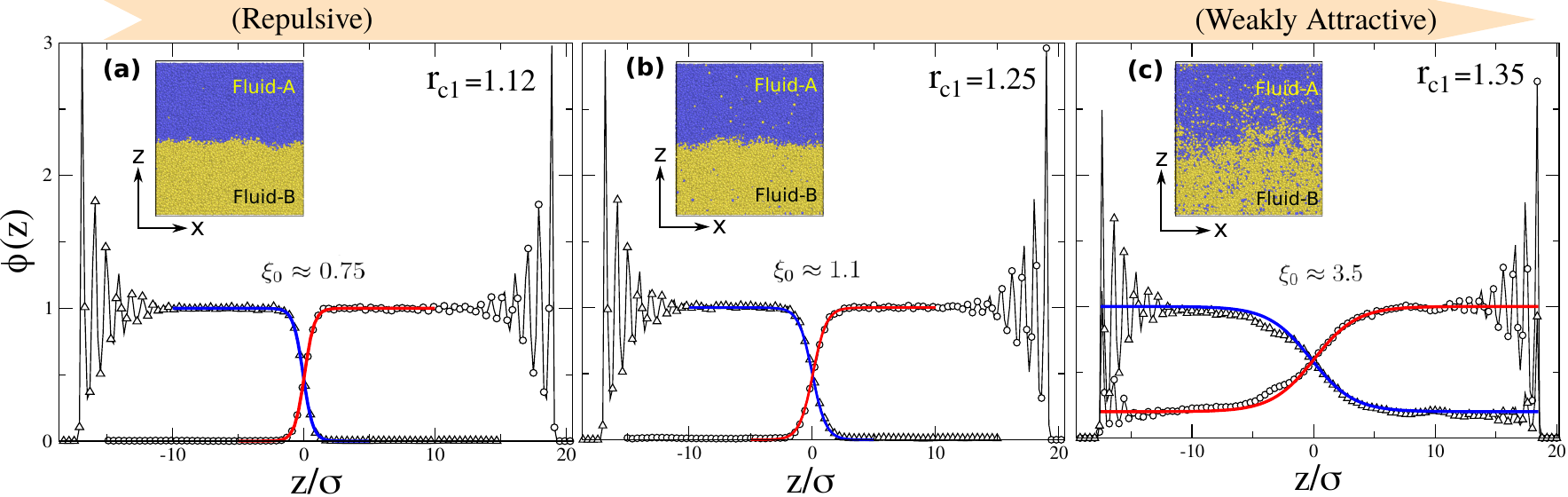}
		\caption{Fluid density $\phi(z)$ (normalized by bulk liquid density) of the confined phase separated pure binary-fluid system is shown for three different values of fluid-fluid interaction cut-off distance $r_{c1}$: (a) 1.12 (repulsive), (b) 1.25, and (c) 1.35 (weakly-attractive) where the system still maintain majority-A and majority-B phases, but a significant amount of inter-diffusion is observed. The mid-point of the interface is located at $z=0$.}
		\label{fig: vol-frac-pure-fluid}
	\end{center} 
\end{figure}

In order to study the effect on the interface width, we first prepare a pure symmetric binary blend (i.e., reference system) at the same density and confined by two 93-LJ wall potential as in the case of system with polymer. Three samples having different interface sharpness (characterized by the interaction cut-off distance in the range: $1.12~({\rm sharp-interface}) \le r_{c1} \le 1.35~({\rm weak-interface})$) are relaxed at $T=1.0$. The laterally averaged density profile $\phi$ (normalized by the bulk density) of the fluid-A/B together with the typical configurations of the system is shown in figure~\ref{fig: vol-frac-pure-fluid}. The density profiles oscillates and the amplitude increases on approaching the walls, a typical behavior near confining walls, also decrease of $\phi$ near the fluid-fluid interface is clearly visible.  

\begin{figure}[ht]
	\begin{center}
		\includegraphics[width=0.9\textwidth]{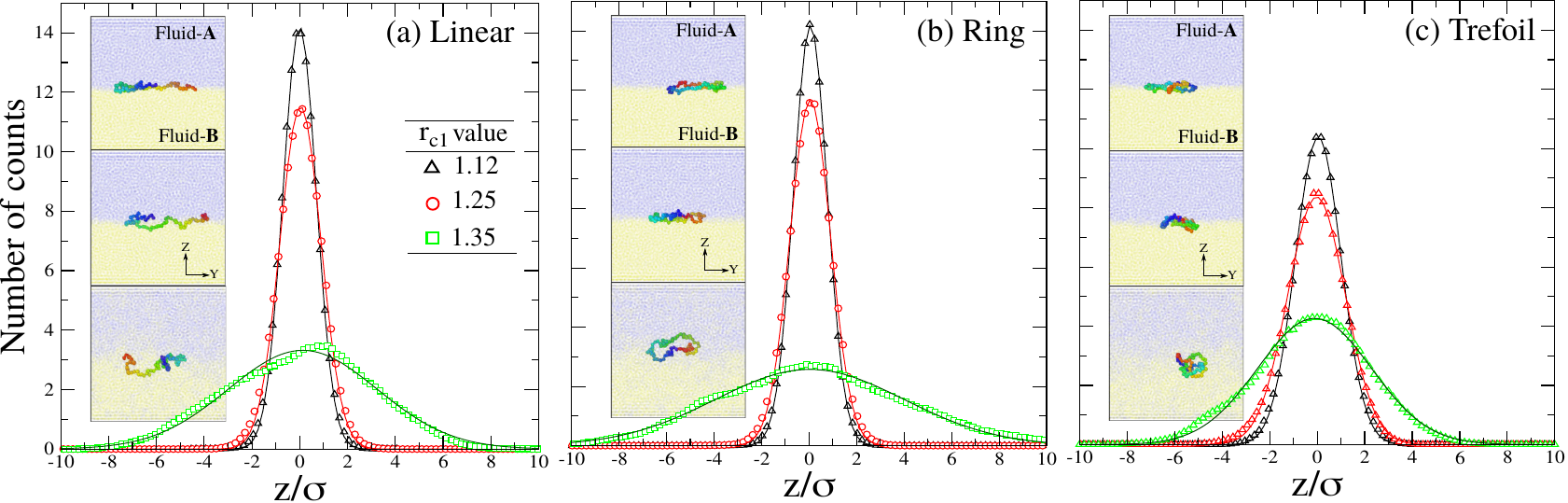}
		\caption{Monomer distribution of (a) Linear, (b) Ring, and (c) Trefoil polymers shown for three different values of fluid-fluid interaction cut-off distance $r_{c1}$: (a) 1.12 (sharp-interface), (b) 1.25, and (c) 1.35 (weak-interface). Solid line represent Gaussian fit to the data. Here, for fluid-polymer interaction $r_{c2}=r_{c}^*~(=1.5)$. Inset figures are the snapshot of the corresponding instantaneous polymer conformation.}
		\label{fig: vol-frac-poly}
	\end{center} 
\end{figure}

For a symmetric binary fluid mixture, the concentration profile across the fluid-fluid interface is given by
\begin{equation}
\phi(z) = \pm \phi_b \tanh\left( \frac{z-z_0}{\xi_0} \right)~,
\label{eqn: interface-profile}
\end{equation} 
where $\phi_b$ represents the concentration in the bulk, $z_0$ marks the mid-point of the interface, and $\xi_0$ is the interface width.\cite{cates2018} The interfacial profile is fixed by a trade-off between the penalty for sharp gradients and the purely local free energy terms. For our binary-fluid systems, the values of interface width are obtained by fitting equation~\ref{eqn: interface-profile} in the density profile near the interface region, see solid lines near the interface in the figure. The values obtained are as follows: $\xi_0 \approx 0.75$ ($r_{c1}=1.12$), $\xi_0 \approx 1.1$ ($r_{c1}=1.25$), and $\xi_0 \approx 3.5$ ($r_{c1}=1.35$). The observation that $\xi_0$ value increases with increasigng $r_{c1}$ is consistent with the fact that as fluid-fluid interaction changes from being repulsive to weakly-attractive the energy barrier of crossing the interface decreases. Thus, the inter-diffusion region at the interface gets larger and consequently the interface width increases. Similarly, for the systems with polymer also we first calculate the density profile (not shown here) and then by fitting equation~\ref{eqn: interface-profile} near the interface region we obtained the values of interface width $\xi$. In addition, we calculate the monomer distribution perpendicular to the interface plane. The monomer distribution obtained at polymer-fluid interaction $r_{c2}=r_{c}^*$ is shown in figure~\ref{fig: vol-frac-poly}. The extent of monomer spread normal to the interface plane is then quantified through the standard deviation (SD) of the Gaussian fit to the monomer density profile. \\

\begin{figure}[ht]
	\begin{center}
		\includegraphics[width=0.8\textwidth]{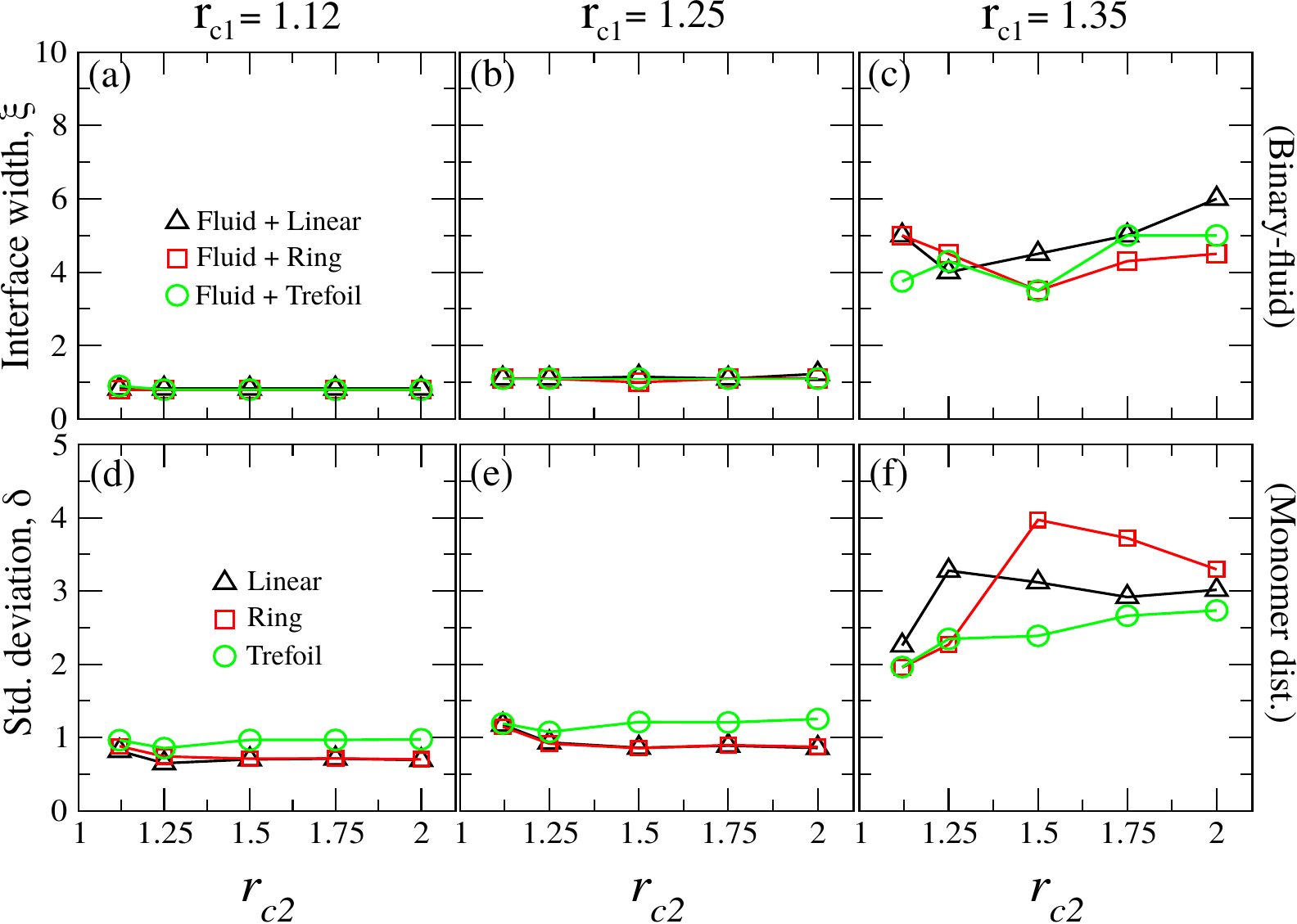}
		\caption{Standard deviation of the monomer distribution shown in figure~\ref{fig: vol-frac-poly} normalized by the pure binary-fluid interface width $\xi_0$ is compared for different topology and at different polymer-fluid interaction strengths. From left to right: (a) sharp-interface, (b) intermediate, and (c) weak-interface.}
		\label{fig: interface-width}
	\end{center} 
\end{figure}

In figure~\ref{fig: interface-width}, we display the variation of fluid interface width $\xi$ and SD of the monomer distribution as a function of $r_{\rm c2}$ for chain in sharp ($r_{c1}=1.12$), intermediate ($r_{c1}=1.25$), and weak ($r_{c1}=1.35$) interface binary-fluid systems. As one can see in figure~\ref{fig: interface-width}(a)-(b), for both sharp and intermediate systems increasing the value of $r_{c2}$ has negligible effect on the interface width with $\xi \approx 0.8$ (sharp) and 1.1 (intermediate) which is almost same as that of the corresponding reference system. 
For the monomer distribution, SD slightly decreases followed by -- (i) remaining constant for linear/ring and (ii) a weak increase for trefoil, with the increase of $r_{c2}$, see figure~\ref{fig: interface-width}(d)-(e). 
On the other hand, for the system having weak interface, the value of $\xi$ decreases and then increases for all the chain types with minimum around $r_{c2}\approx1.25$ (linear) and $r_{c2}\approx1.5$ (ring and trefoil). 
The interface width varies roughly in the range $4-6$ (linear) and $3.5-5$ (ring and trefoil), see figure~\ref{fig: interface-width}(c). 
Whereas, SD of the monomer distribution increases and then decreases for linear and ring with the maximum occuring at the positions where $\xi$ attains minimum, and for trefoil SD increases gradually. The value of SD varies roughly in the range $2.2-3.3$ (linear), $1.9-4$ (ring), $1.9-2.7$ (trefoil). \\
Thus, from the above observations it is clear that when the fluid-fluid interface is relatively sharp adding a chain (irrespective of topology and polymer-fluid interaction strength) produce negligible effect on the fluid interface width. However, a significant effect may be observed with much higher chain concentration (or larger chain) at the interface. On the other hand, the extent of monomer spread (normal to the interface plane) depends on the chain topology. To gain further insight we study the chain size and their instantaneous shape upon chainging the parameters in the following section.


\subsection{Chain size and instantaneous shape}  
\label{sec: chain size and shape}

The chain extension is characterized by the mean-square radius of gyration and it is defined as 
\begin{equation} 
\left\langle R^2_g \right\rangle = \left\langle \frac{1}{N}\sum_{i=1}^N({\bf r}_i-{\bf r}_{_{cm}})^2\right\rangle~,
\end{equation} 
with $N$ the number of monomers per chain, ${\bf r}_i$ and ${\bf r}_{_{cm}}$ the monomer position and center-of-mass of the chain, respectively, and angular bracket represents the ensemble average. Since the chain is located at the fluid-fluid interface the chain extension parallel, $R_{g(\parallel)} = \left< R^2_g(x,y) \right>$, and perpendicular, $R_{g(\perp)} =  \left< R^2_g(z) \right>$, to the interface plane are also calculated.   

\begin{figure}[ht]
	\begin{center}
		\includegraphics[width=0.9\textwidth]{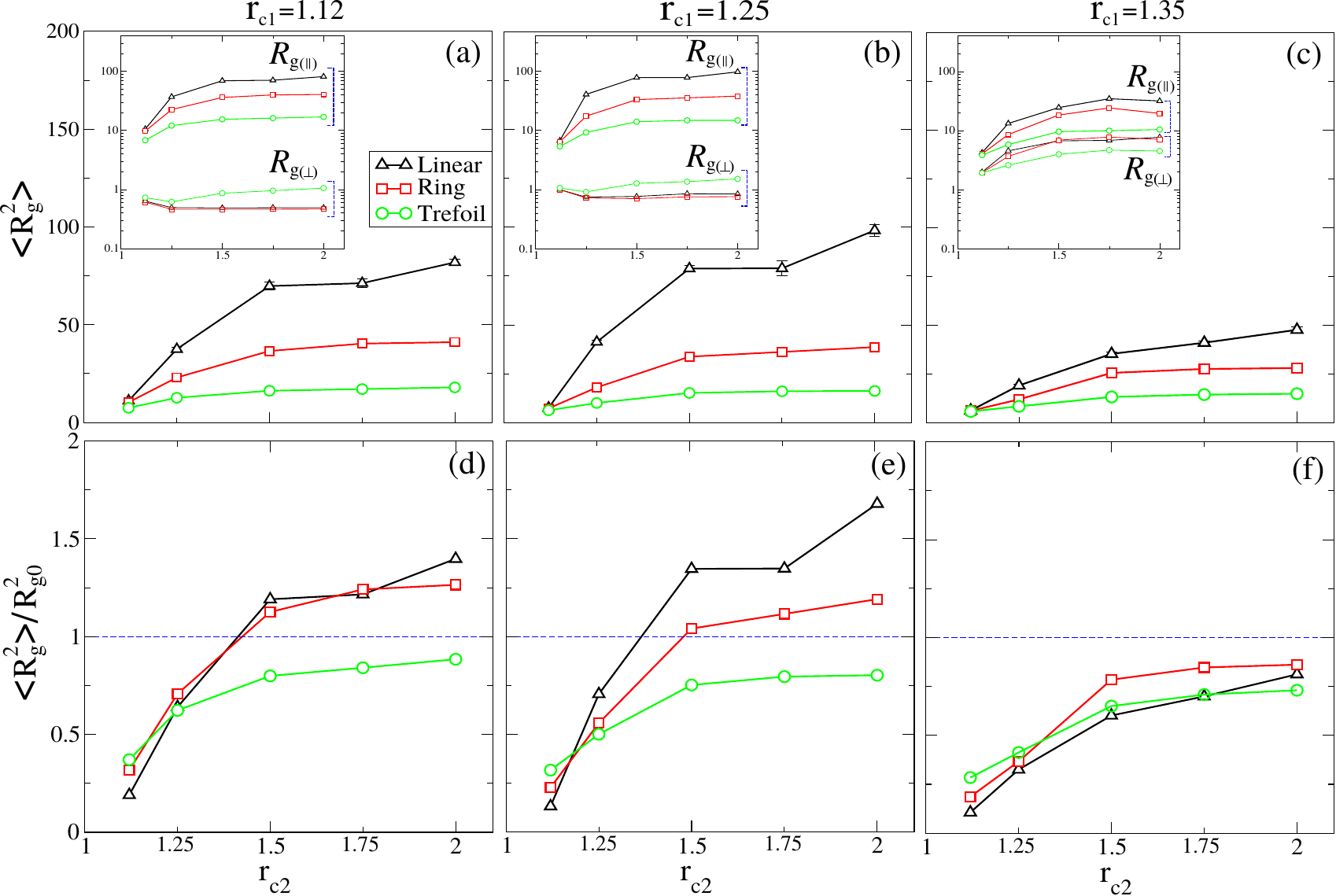} 
		\caption{{\bf Upper panel-} $\left< R_g^2 \right>$ is plotted along with the corresponding parallel, $R_{g(\parallel)}$, and perpendicular, $R_{g(\perp)}$, components (inset figure) as a function of fluid-polymer interaction cut-off $r_{c2}$ shown for three different values of fluid-fluid interaction cut-off $r_{c1}$: (a) 1.12 (sharp-interface), (b) 1.25, and (c) 1.35 (weak-interface). {\bf Lower panel-} Radius of gyration normalized by the chain size in the absence of explicit fluid particles. 
		}
		\label{fig: rg}
	\end{center} 
\end{figure}

In figure~\ref{fig: rg}(a)-(c), we display the mean-square radius of gyration together with the parallel and perpendicular components as a function of increasing fluid-polymer interactions $r_{c2}$. It is observed that as we increase the polymer-fluid interaction range, $r_{\rm c2}$, $\left< R_g^2 \right>$ increases and reaches maximum at $r_{c2}\approx1.5$ ($=r_{c}^*$, self-interaction of the fluid) for all chain types and then saturates for ring and trefoil, whereas for linear chain it remains constant until $r_{c2}\approx1.75$ followed by a slight increase again, see figure~\ref{fig: rg}(a)-(b). Also for the weak interface system, see figure~\ref{fig: rg}(c), the chain size increases continusly for linear, while it saturates for both ring and trefoil. In all the considered parameter values (except when $r_{c2}=1.12$) the average chain size is consistently largest for linear followed by ring and trefoil. The effect of topological constraint is clearly visible if we compare the parallel and perpendicular components. 
In the case of both sharp and intermediate interface binary-fluid systems, $R_{g(\parallel)}$ increases and then saturates for all the three chain types, while $R_{g(\perp)}$ decreases and saturaes for both linear and ring. In contrast, for trefoil $R_{g(\perp)}$ decreases slightly and then increases. On the other hand, for system with weak interface both parallel and perpendicular components increases in a similar way. Furthermore, we also calculate the relative chain size $\left< R_g^2 \right>/R_{g0}^2$, with $R_{g0}^2$ the size of corresponding athermal chain (i.e., polymer in implicit solvent and monomer-monomer interaction cut-off at LJ potential minimum), see figure~\ref{fig: rg}(d)-(f). For the system with sharp interface, see figure~\ref{fig: rg}(d), the ratio increases and approaches unity when $r_{c2}\approx 1.4$ for both linear and ring, while for trefoil the ratio is always less than unity (even for intermediate and weak interface systems). For intermediate sharpness, see figure~\ref{fig: rg}(e), similar trend is observed; however, in comparison the linear (ring) chain size is larger (smaller), and the ratio attains unity at $r_{c2}\approx$ 1.36 (linear), 1.48 (ring). Interestingly, for the weak interface system, see figure~\ref{fig: rg}(f), although all the chains show increasing trend the ratio never reaches unity.  


\begin{figure}[ht]
	\begin{center}
		\includegraphics[width=0.9\textwidth]{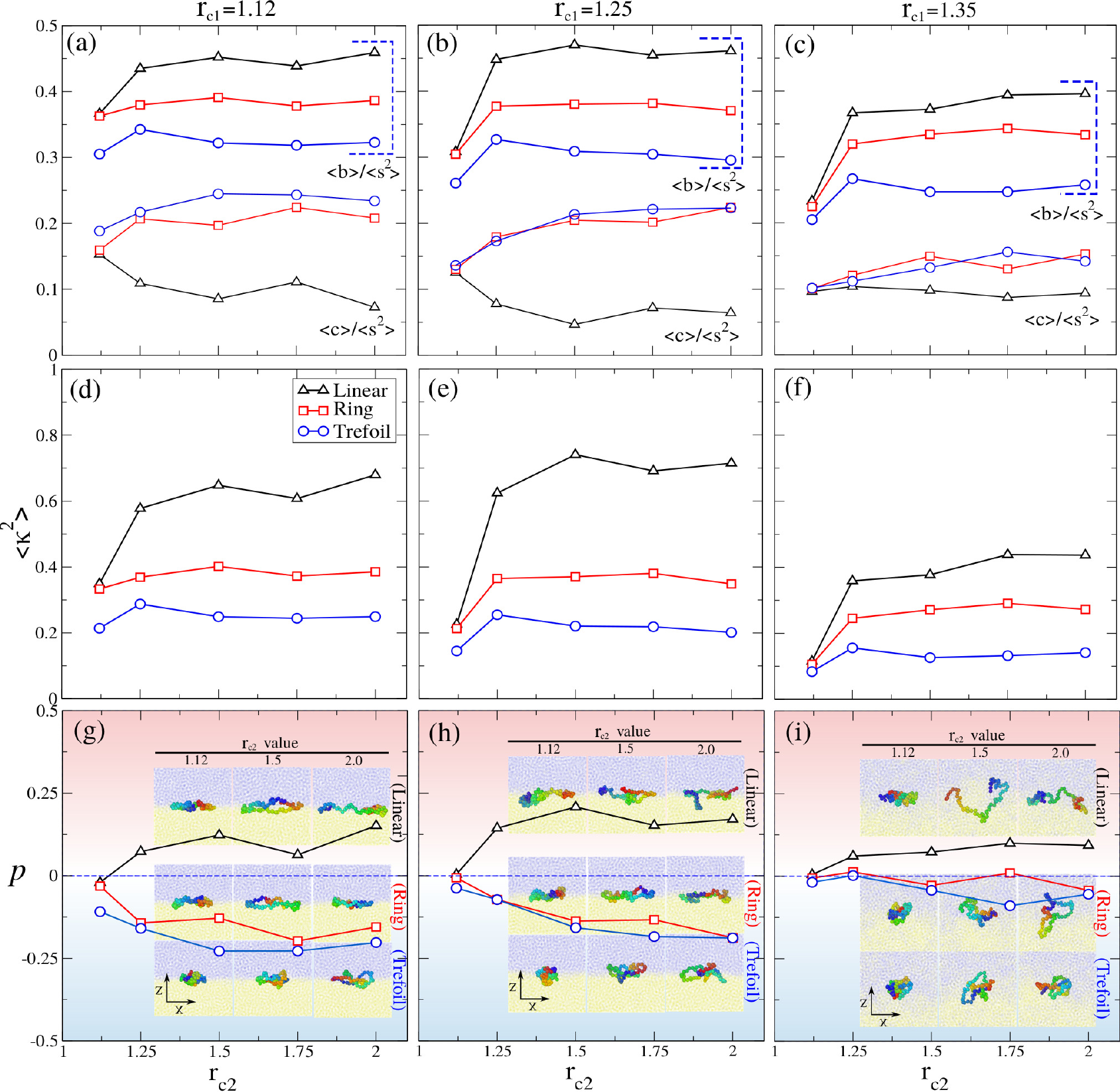}
		\caption{Normalized shape parameters $\langle b \rangle / \langle s^2 \rangle$ and $\langle c \rangle / \langle s^2 \rangle$ (upper panel), $\kappa^2$ (middle panel), and $\left< p \right>$ (lower panel) of linear, ring, and trefoil chains are plotted as a function of fluid-polymer attraction cut-off distance $r_{c2}$ (with stronger attraction for larger $r_{c2}$ values) shown for binary-fluid with sharp ($r_{c1}$=1.12), intermediate ($r_{c1}$=1.25) and weak ($r_{c1}$=1.35) interfaces as indicated at the top of the figure. Lower panel inset figure: Typical snapshot of the chains at the indicated value of $r_{\rm c2}$.}  
		\label{fig: shape-param-1}
	\end{center} 
\end{figure}

It is known that the average instantaneous shape of a polymer chain is not spherical in contrast to the orientation averaging picture. For example, the average instantaneous shape of a linear chain is an ellipsoid rather than a sphere. The instantaneous shape is studied by means of the radius of gyration tensor {\bf S} as a shape measure.\cite{theodorou1985,solc_stockmayer_1971, solc1971, mattice_book} By choosing a principal axis system where {\bf S} is in diagonal form and the eigenvalues $\lambda_1^2, \lambda_2^2$ and $\lambda_3^2$ are such that $\lambda_1^2 \le \lambda_2^2 \le \lambda_3^2$. The shape anisotropy of a polymer chain is characterized in terms of the following parameters   
\begin{align}
b &= \lambda_3^2 - \frac{1}{2} (\lambda_1^2+\lambda_2^2)~,\nonumber \\
c &= \lambda_2^2 - \lambda_1^2~, ~\text{and} \nonumber \\
\kappa^2 &= (b^2+\frac{3}{4}c^2)/s^4~,
\label{eqn:shape-param}
\end{align}
\noindent where $b$, $c$, and $\kappa^2$ are asphericity, acylindricity, and relative shape anisotropy, respectively, and $s^2=\lambda_1^2+\lambda_2^2+\lambda_3^2 = \text{tr}({\bf S})$ is the squared radius of gyration of the chain. Here, the quantity $b$ and $c$ measure the deviations from spherical and cylindrical shapes, respectively, with $b=0$ and $c=0$ for perfect sphere and cylinder, respectively. Also, we calculate the prolateness parameter $p$ defined as   
\begin{equation}
p = \frac{(2\lambda_1-\lambda_2-\lambda_3)(2\lambda_2-\lambda_1-\lambda_3)(2\lambda_3-\lambda_1-\lambda_2)}{2(\lambda_1^2+\lambda_2^2+\lambda_3^2-\lambda_1\lambda_2-\lambda_2\lambda_3-\lambda_3\lambda_1)^{3/2}}~,
\label{eq: prolateness}
\end{equation} 
where the prolateness parameter $p=1$ for perfectly prolate shape and $p=-1$ for perfectly oblate shape. \\

We calculate the shape parameters defined above for all the chain types and plotted as a function of $r_{c2}$ considering the three different binary-fluid systems, see figure~\ref{fig: shape-param-1}. With the increase of $r_{c2}$ the value of $\left< b \right>$ increases for all the chain types implying increasing deviation from the spherical shape, while $\left< c \right>$ decreases for linear (i.e., becoming more cylindrical) and increases for ring and trefoil. In comparison, the value of asphericity (acylindricity) is highest (lowest) for linear chain followed by ring and trefoil in the entire range of $r_{c2}$ (except at $r_{c2}=1.12$, where linear and ring have roughly the same value), see figure~\ref{fig: shape-param-1}(a)-(c). Note that the acylindricity curves for ring and trefoil closely follow each other. These observation clearly suggest that the linear chain is highly cylindrical, whereas ring and trefoil are relatively more spherical. Furthermore, the value of $\left< \kappa^2 \right>$ is rougly equal for linear and ring at $r_{c2}=1.12$ and the value is smaller for trefoil (for sharp and intermediate interface binary-fluid). While for other values of $r_{c2}$ it is highest for linear followed by ring and trefoil, see figure~\ref{fig: shape-param-1}(d)-(f). Finally, in figure~\ref{fig: shape-param-1}(g)-(i) we show the averaged prolateness parameter $\left< p \right>$ as a function of $r_{c2}$. It is observed that, when the binary-fluid has sharp interface the value of $\left< p \right>$ for linear polymer can vary from negative ($\approx -0.02$) at $r_{c2}=1.12$ to positive for the other higher values of $r_{c2}$ considered, see figure~\ref{fig: shape-param-1}(g), implying a shape transition from oblate to prolate. Whereas, for ring and trefoil, the value of $\left< p \right>$ is always negative and show a decreasing trend with increasing $r_{c2}$. Similar trend is observed for chains in the binary fluid system with $r_{c1}=1.25$ (intermediate). For the chains in weak-interface fluid system and the value of $\left< p \right>$ moves closer to zero from above (below) for linear (ring and trefoil); however, for ring we also see that $\left< p \right>$ becomes slightly positive ($\approx 0.01$) at $r_{c2}=1.25,~1.75$, see figure~\ref{fig: shape-param-1}(i).  



\subsection{Chain dynamics at the interface}
\label{sec: dynamics} 

\begin{figure}[ht]
	\begin{center}
		\includegraphics[width=0.8\textwidth]{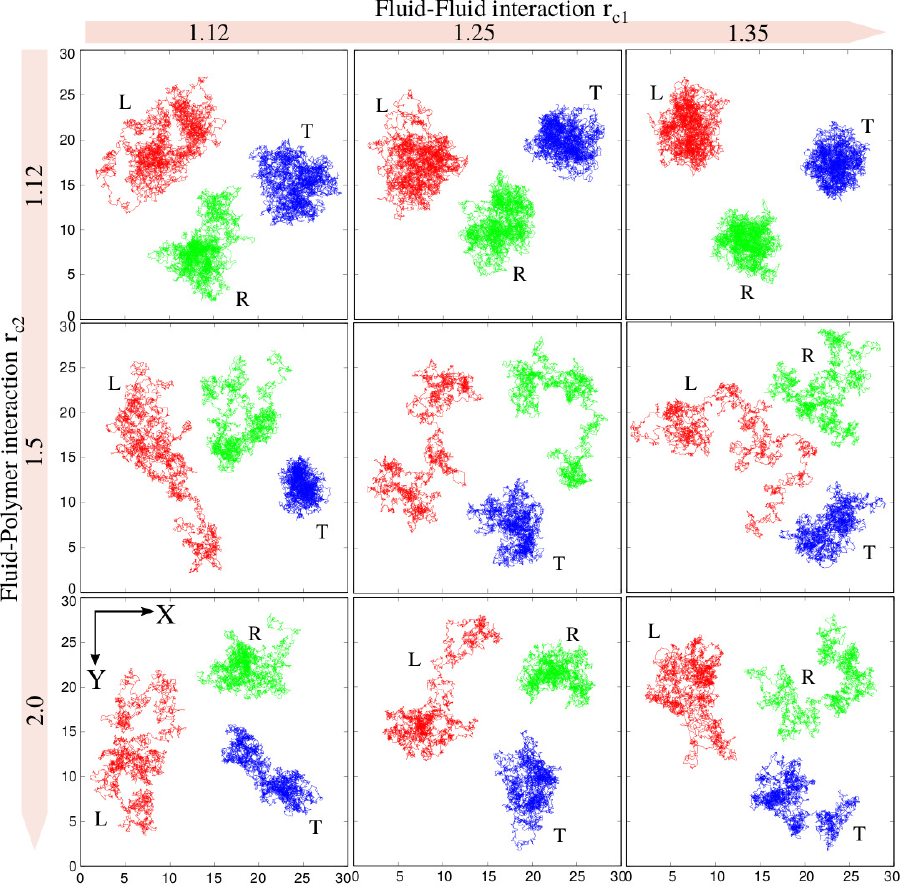}
		\caption{Trajectories of the center-of-mass of linear (L), ring (R), and trefoil (T) chains traced out during a simulation run of 15 million MD steps are shown for different values of fluid-fluid interaction (characterized by cut-off value $r_{c1}$) and fluid-polymer interaction (characterized by cut-off value $r_{c2}$) indicated in the figure. In the figure, going from left to right the fluid-fluid interface changes from being sharp to broad, and from top to bottom polymer-fluid interaction changes from being repulsive to attractive.}
		\label{fig: com trajectories}
	\end{center} 
\end{figure}

We have seen that the linear, ring, and trefoil chains respond to the environment differently, i.e., variation of size and shape upon changing fluid-fluid/polymer interactions are different and thus may affect the chain dynamics at the interface. For example, as shown in figure~\ref{fig: com trajectories}, the area traced out by the center of mass trajectory in the interface plane is different for diffrent chain types of same molecular weight. Also note that for the sharp interface binary-fluid system ($r_{c1}=1.12$ in the figure) with increasing fluid-polymer interaction (i.e., $r_{c2}=1.12/{\rm repulsive}-2.0/{\rm attractive}$) the area traced out by the trefoil shrinks followed by an increase again, whereas it is different for linear and ring polymers. Therefore, to quantify these trajectories and compare among the chains we calculate mean-square displacement (MSD) for different parameter values.  

On the other hand, understanding the behavior of diffusion at the interface is important because of its relevance to various practical applications such as in adhesion, coating, lubrication, micro-/nano-fluidics, etc.~and biological mass transporting processes.\cite{granick2003}
In this study, we focus the chain motion along the interface plane only since the chain is confined to the fluid-fluid interface and thus the motion perpendicular to the plane is restricted. 
Furthermore, since only one chain is present at the interface we calculate the time averaged MSD (instead of ensemble averaged) as done in single particle tracking experiments where the quantity is extracted from the available position time series ${\bf r}(t)$ of a tracer molecule. Mathematically, the time averaged MSD is defined as  

\begin{equation}
\overline{\delta^2(\Delta,t)} \equiv \frac{1}{t-\Delta} \int_{0}^{t-\Delta} \left[{\bf r}(t'+\Delta)-{\bf r}(t')\right]^2 dt'~,
\label{eqn: msd} 
\end{equation}
where the overline $\overline{(.)}$ denotes the time average. Here, $\Delta$ is the lag time constituting a time window swept along the time series which is less than the total measurement time $t$. The time averaged MSD thus compares the particle positions along the trajectory which is separated by the time difference $\Delta$. Over long-time measurement, the time averaged MSD is related to the diffusion exponent $\alpha$ as 
\begin{equation}
\overline{\delta^2} \sim D_0\Delta^{\alpha}~,
\end{equation}
where $D_0$ is the diffusion constant and the exponent typically lies in the range $0 < \alpha \le 1$ (with $\alpha=1$ for Brownian motion and $\alpha < 1$ for subdiffusive or anomalous diffusion).  
Since the system under study is in equilibrium (fluid phase) the time averaged quantity from a sufficiently long time series should be equal to the corresponding ensemble average.\cite{vanKampen,khinchin} Hereafter, we use MSD to denote time averaged MSD (otherwise stated clearly). To this end we calculate the following MSD quantities:
\begin{equation}
g_1(\Delta,t) = \left< \frac{1}{t-\Delta} \int_{0}^{t-\Delta} \left[{\bf r}_i(t'+\Delta)-{\bf r}_i(t')\right]^2 dt' \right> ~,
\label{eqn: msd-g1}
\end{equation}
which describes the monomeric MSD with ${\bf r}_i$ denoting the position of $i^{\rm th}$ monomer, and $\left< \cdot \right>$ denotes average over all the monomers in a chain. Thus, $g_1(\Delta,t)$ represents the ensemble average of the time averaged MSD. The above quantity with repsect to the center-of-mass (CM) position defined as 
\begin{equation}
g_2(\Delta,t) = \left< \frac{1}{t-\Delta} \int_{0}^{t-\Delta} \left( \left[{\bf r}_i(t'+\Delta) - {\bf r}_{_{\rm CM}}(t'+\Delta)\right]^2 -\left[{\bf r}_i(t') - {\bf r}_{_{\rm CM}}(t') \right]^2 \right) dt' \right> ~.
\label{eqn: msd-g2}
\end{equation} 
And finally MSD for the CM position of the chain defined as
\begin{equation}
g_3(\Delta,t) = \frac{1}{t-\Delta} \int_{0}^{t-\Delta} \left[{\bf r}_{_{\rm CM}}(t'+\Delta)-{\bf r}_{_{\rm CM}}(t')\right]^2 dt' ~, 
\label{eqn: msd-g3}
\end{equation}
where ${\bf r}_{_{\rm CM}}$ denotes the position of the CM.  

\begin{figure}[ht]
	\begin{center}
		\includegraphics[width=0.85\textwidth]{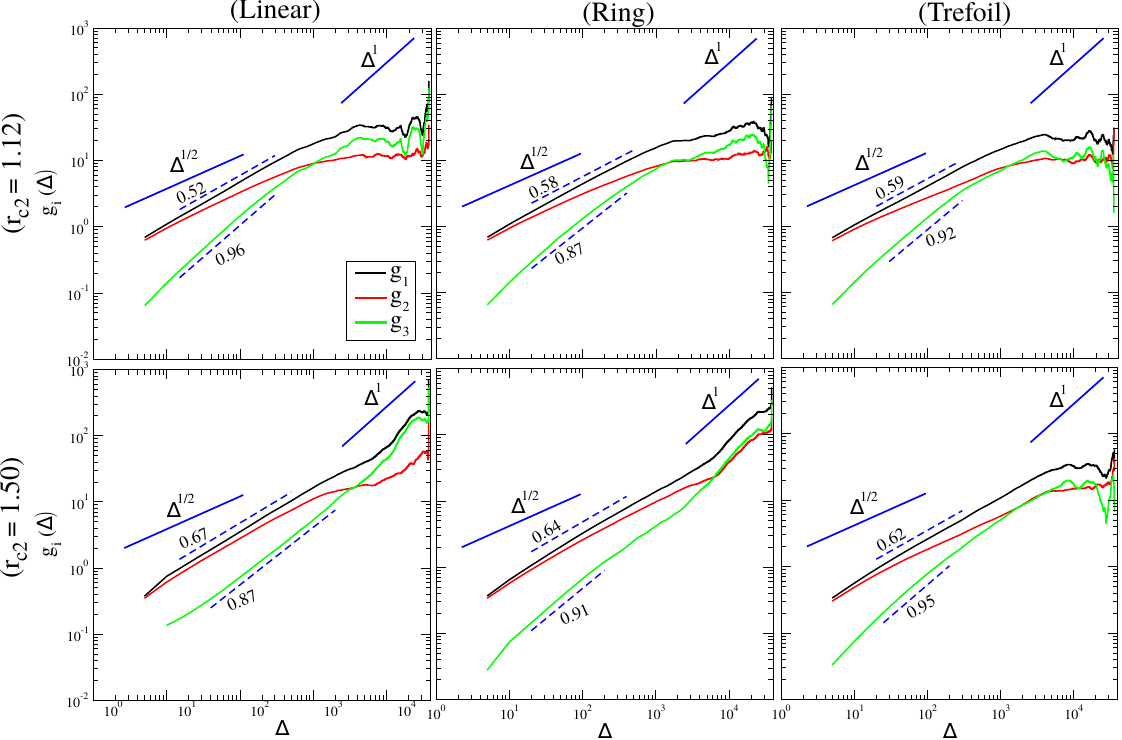}
		\caption{Mean square displacements $g_1$, $g_2$, and $g_3$ for polymer (linear, ring, and trefoil) at the fluid-fluid interface with intermediate sharpness ($r_{c1}=1.25$) when the polymer-fluid interaction is repulsive ($r_{c2}$ = 1.12, upper panel) and attractive ($r_{c2}$ = 1.5, lower panel). The scaling laws $\Delta^{1/2}$ and $\Delta^{1}$ are also indicated in the figure.}
		\label{fig: msd-g1g2g3}
	\end{center} 
\end{figure}

In figure~\ref{fig: msd-g1g2g3}, we display the MSDs $g_1$, $g_2$, and $g_3$ for polymer at the binary-fluid interface with intermediate sharpness shown for repulsive ($r_{c2}=1.12$) and weakly-attractive ($r_{c2}=r_c^*=1.5$) polymer-fluid interaction cases. Here, monomer positions are measured with respect to the CM of the system. In both cases, it is observed that $g_1 > g_3$ at short to intermediate timescales, and as expected for the polymeric system, $g_1$ shows sub-diffusion with exponent $\alpha$ in the range $0.52-0.59$ (for $r_{\rm c2}< r_{c}^*$, i.e., repulsive polymer-fluid interaction relative to self-interaction of the fluid) and $0.62-0.67$ (for $r_{\rm c2} \ge r_{c}^*$, i.e., attractive fluid-polymer interaction relative to self-interaction of the fluid) systems. While the value of exponent for $g_3$ is much closer to 1 with value in the range $0.87-0.96$. As one can see in the figure, expected crossover to diffusive regime for $g_1$ and $g_3$ is not observed for some of the parameter values within the reported simulation time scale and therefore requires a much longer simulation runs. In the following, we restrict our discussions upto intermediate time scales.  

\begin{figure}[ht]
	\begin{center}
		\includegraphics[width=0.9\textwidth]{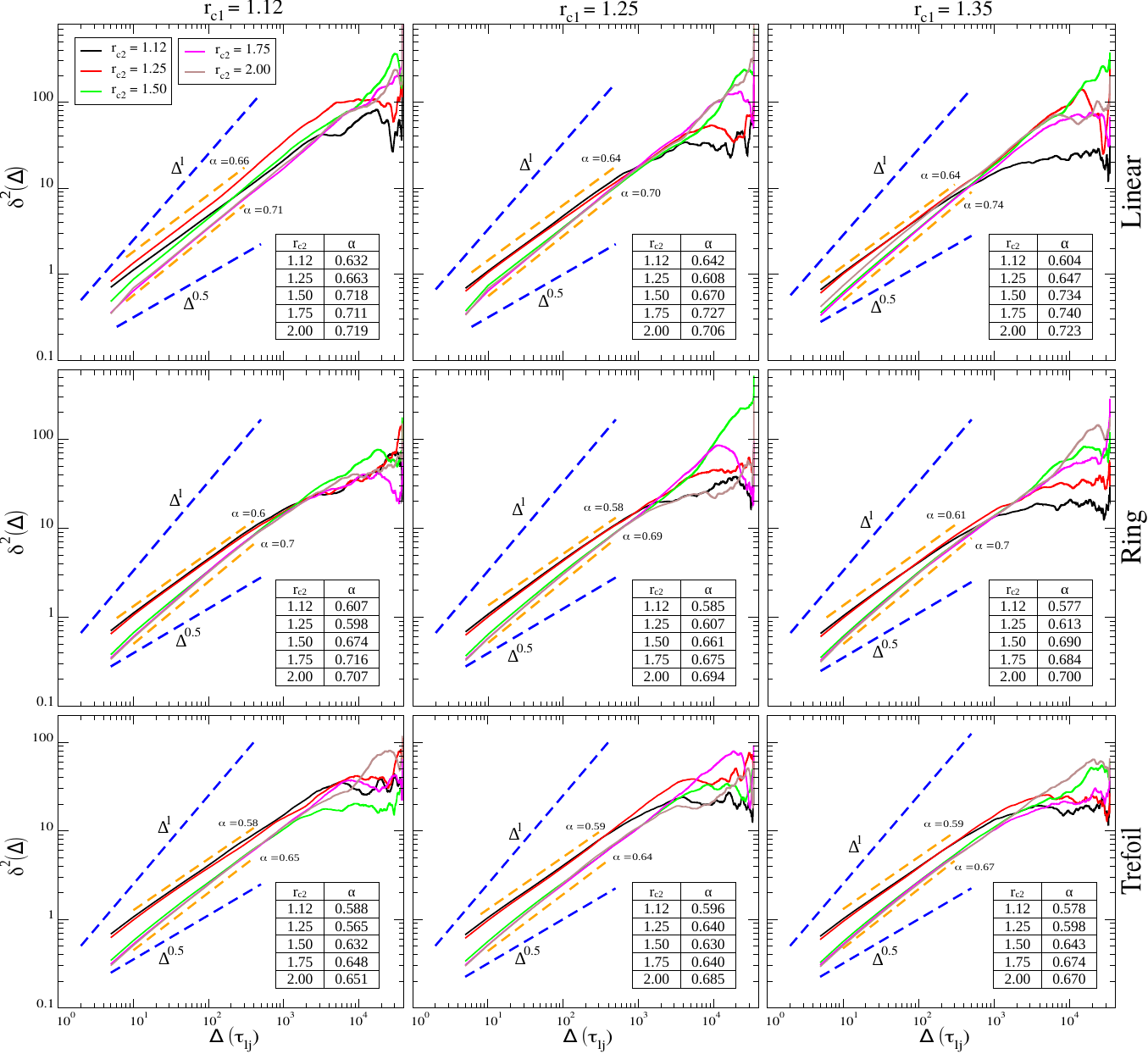}
		\caption{Monomeric MSD, $g_1(\Delta)$, for linear (top panel), ring (middle panel), and trefoil (lower panel) chains at five different values of polymer-fluid interaction indicated by $r_{c2}$ in the figure are shown for $r_{c1}=$ 1.12 (sharp interface), 1.25 (intermediate), 1.35 (weak interface). Dashed lines with $\delta^2\sim \Delta$ and $\delta^2\sim \Delta^{0.5}$ is guide to the eye. Inset table indicates the value of exponent (or slope) $\alpha$ obtained in the linear region.}
		\label{fig: msd-monomeric}
	\end{center} 
\end{figure}

\begin{figure}[ht]
	\begin{center}
		\includegraphics[width=0.9\textwidth]{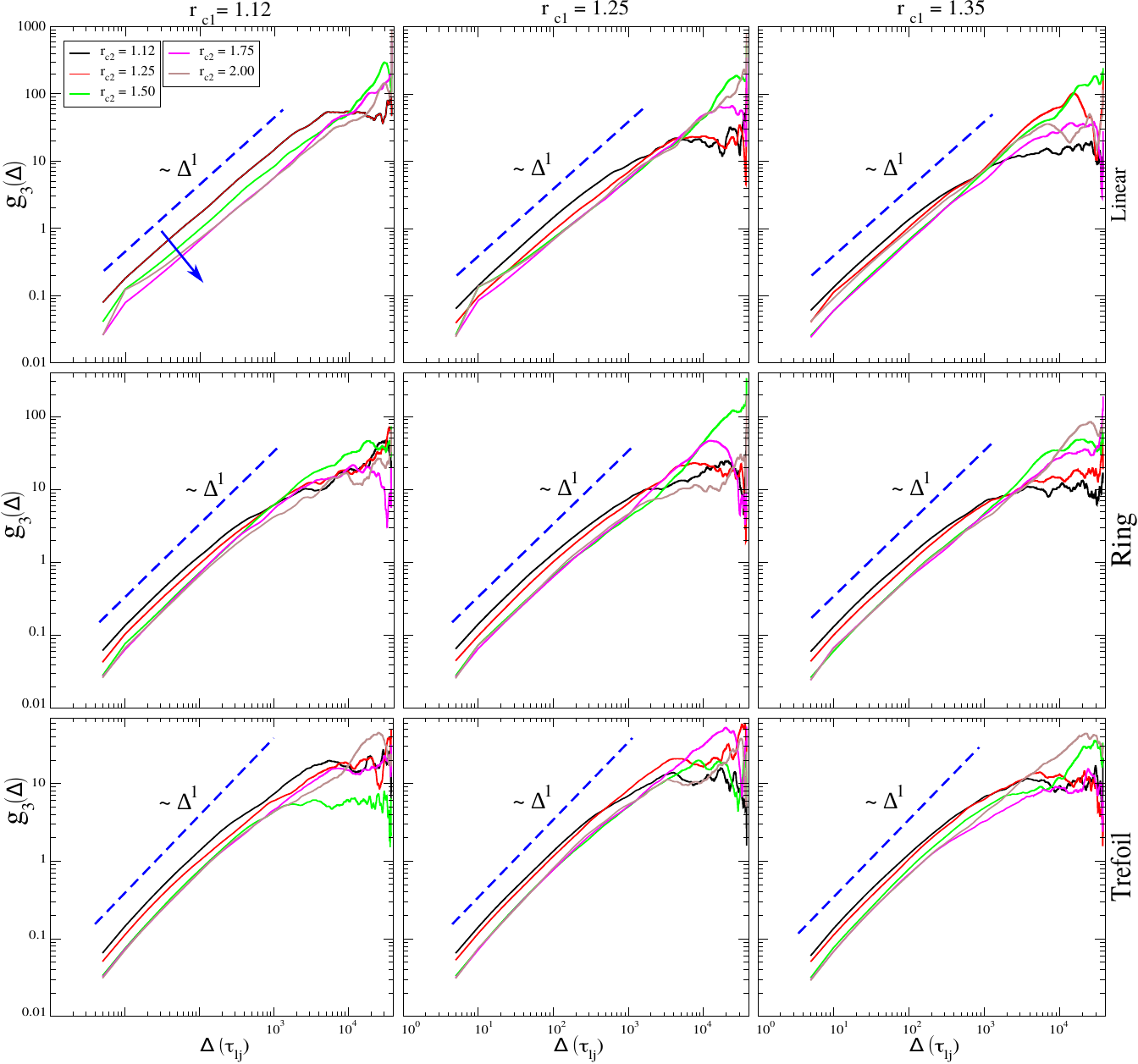}
		\caption{Center-of-mass MSD, $g_3(\Delta)$, curves for linear (top panel), ring (middle panel), and trefoil (lower panel) at different values of polymer-fluid interaction range $r_{c2}$ are shown for $r_{\rm c1}$=1.12 (sharp interface), 1.25 (intermediate), 1.35 (weak interface) indicated at the top of the figure. Dashed line with $g_3(\Delta)\sim\Delta$ is guide to the eye, and downward arrow at the top-left figure indicates the direction of increasing $r_{\rm c2}$ which is same for all the figures.}
		\label{fig: msd-com}
	\end{center} 
\end{figure}

\begin{figure}[ht]
	\begin{center}
		\includegraphics[width=0.7\textwidth]{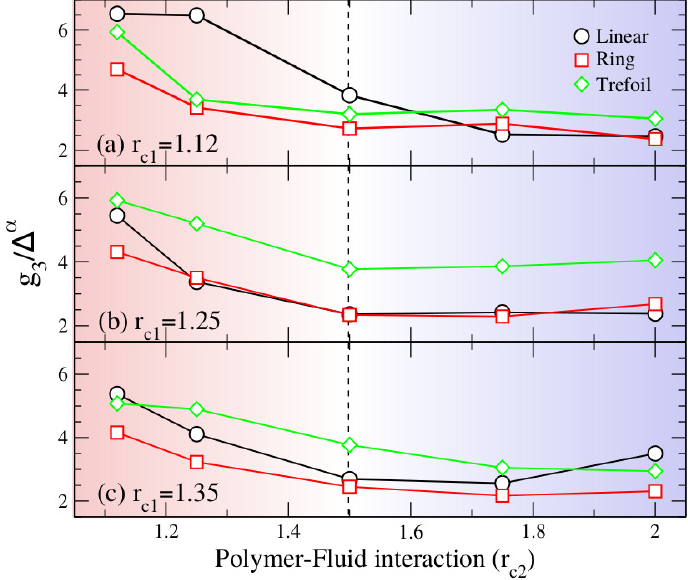}
		\caption{Plot of $g_3/\Delta^\alpha$ as a function of polymer-fluid interaction range/value, $r_{\rm c2}$, for linear, ring, and trefoil chains shown for binary-fluid system with (a) $r_{\rm c1}$=1.12 (sharp-interface), (b) $r_{\rm c1}$=1.25 (intermediate), and (c) $r_{\rm c1}$=1.35 (weak-interface). Verical dashed line at $r_{\rm c2}$=1.5 seperates the left and right regions, where on the left (or right) the polymer-fluid attraction is less (or more) than the self-interaction of pure fluid phase.}
		\label{fig: diff-coeff}
	\end{center} 
\end{figure}

In figure~\ref{fig: msd-monomeric} and~\ref{fig: msd-com} we display the MSD curves $g_1$ and $g_3$, respectively, for all the chain types at different values of polymer-fluid interaction range, $r_{\rm c2}$, shown for chains in the considered three binary-fluid systems of different interface sharpness. It is interesting to note that, $g_1$ (and $g_2$, see supplementary figure) irrespective of chain types and interface sharpness, the curves are grouped into two with each group having roughly the same value of exponent $\alpha$. As indicated in figure~\ref{fig: msd-monomeric}, the value of $\alpha$ is smaller (or larger) for $r_{\rm c2} < r_{c}^*$ (or $r_{\rm c2} \ge r_{c}^*$) suggesting that there is qualitative change in the early time dynamical behavior when the fluid-polymer interaction cut-off distance, $r_{\rm c2}$, equals that of the corresponding self-interaction of the fluid particles, $r_{c}^*$, see table~\ref{table: cut-off-radii}. This qualitative change may be related to the conformational changes, i.e., for $r_{\rm c2} < r_{c}^*$ the fluid-polymer interaction nature is repulsive w.r.t.~the self-interaction of the fluid-particles and hence to polymer the fluid acts as poor solvent and consequently the chain is confined to the interface (quasi-2D chain), while for $r_{\rm c2} \ge r_{c}^*$ the fluid acts as good solvent and as a result chain adopts a more extended shape (3D chain). Furthermore, in changing the conformation from quasi-2D to 3D the chain protrude deeper in the pure fluid phases and experience larger drag force as indicated by the smaller value of MSD. It is interesting to note that, for the considered chain length, the observed qualitative change is independent of the chain topology. For the chain CM motion, we observe that $g_3\sim \Delta^\alpha$ with $\alpha\sim 1$ for all the considered parameter values and chain types. 
And from the linear growth region we estimate the diffusion constant, $D_0 \approx g_3/\Delta^\alpha$, and compare among the chain types, see figure~\ref{fig: diff-coeff}. Note that for the chain confined to the sharp-interface ($r_{\rm c1}$=1.12 in the figure) the value of $D_0$ is largest for linear followed by trefoil and ring when $r_{\rm c2} \le r_{\rm c}^*$, while for $r_{\rm c2} > r_{\rm c}^*$ it is largest for trefoil. For the case of intermediate value of interface sharpness trefoil knot chain has the largest diffusion constant for all values of $r_{\rm c2}$, and both ring and linear have almost the same value in the entire range. However, for the case of weak-interface binary-fluid system, $D_0$ is largest for trefoil (except when $r_{\rm c2}$=1.12,~2.0, where linear chain has larger value) followed by linear and ring. It is interesting to note that ring, despite having a smaller size relative to linear, diffuses slower in most of the cases and thus reflects the consequence of chain topology in lateral motion at the fluid-fluid interface.  


\section{Conclusion}
\label{sec: conclusion}

We have presented a comprehensive study on the static and dynamic behavior of a single chain located at the binary-fluid interface as a function of changing polymer-fluid interaction nature, i.e., from repulsive to attractive, and further investigated its effect on the fluid interface width. In particular, we have explored the role of chain topology on the aforesaid properties by considering three topologically different polymer chains, namely, linear, ring and trefoil.\\

For relatively sharp-interface binary-fluid systems, it is observed that the fluid interface width is insensitive to the chain topology and fluid-polymer interaction nature, i.e., repulsive or attractive. However, for the weak-interface system changing polymer-fluid interaction from being repulsive to attractive produces a non-monotonic change of interface width (i.e.~decreases and then increases). For linear (or ring/trefoil) chain the minimum is attained at the polymer-fluid interaction value, $r_{\rm c2}$, less than (or equal to) the self-inteaction value of the pure fluid phase, $r_{\rm c}^*$.  
On the other hand, from the analysis of chain extension normal (or parallel) to the interface plane through radius of gyration it is found that trefoil, due to the presence of knot, extention in the normal (or parallel) direction is relatively large (or small) compared to ring and linear chains and thus highlight the effect of topological constraint on the extension behavior at the interface which may have relevance to practical applications. Furthermore, analysis of the average instantaneous shape parameters revel that chain with trefoil knot is relatively more spherical in shape with least relative shape anisotropy value. And from the study of prolateness parameter $p$ it is also found that linear chain can adopt oblate shape ($p\le0$) under strong confinement at the fluid interface (i.e., when $r_{\rm c2}=1.12$ or repulsive) which upon changing the nature of polymer-fluid interaction to attractive transforms to prolate shape. Whereas, ring and trefoil chains take prolate shape ($p<0$) in almost all the considered parameter range (and $p\approx0$ for some values of $r_{\rm c2}$ for ring in the weak-interface binary-fluid system). \\

Finally, to gain insight on the interplay between the chain topology and dynamics, we have studied the chain motion at the interface through time-averaged MSDs ($g_1,~g_2$, and $g_3$). We observed that both $g_1$ and $g_2$ show sub-diffusive behavior in the early to intermediate time scales. However, irrespective of the chain topology and binary-fluid type, the MSD curves (of $g_1$ and $g_2$) are grouped into two depending on the value of polymer-fluid interaction $r_{\rm c2}$. It was observed that for $r_{\rm c2} < r_{\rm c}^*$ (where fluid acts as a poor solvent to the polymer) the diffusion exponent $\alpha$ is smaller and larger for $r_{\rm c2} \ge r_{\rm c}^*$ (where fluid acts as good solvent and chain swells). Thus, the abrupt change in the value of exponent $\alpha$ correspond to the confomational change, i.e., strongly confined quasi-2D chain ($r_{\rm c2} < r_{\rm c}^*$) to 3D chain ($r_{\rm c2} \ge r_{\rm c}^*$). Whereas, $g_3$ (center-of-mass MSD) show diffusion with exponent $\alpha\approx 1$ for all types of chain topology, interface sharpness, and $r_{\rm c2}$ values. And from the comparison of diffusion constant $D_0\approx g_3/\Delta^\alpha$ we observed that trefoil knot chain has the largest diffusion constant in all the parameter range (except for sharp-interface binary-fluid with $r_{\rm c2} < r_{\rm c}^*$, where linear chain diffuses faster than both ring and trefoil). However, it would also be necessary to explore the chain length dependence of self-diffusion constant in order to fully understand the dynamics at the fluid-fluid interface.


\begin{acknowledgements}
Helpful comments from T.~Kreer and M.~Lang is greatfully acknowledged.
\end{acknowledgements}


\end{document}